\documentclass[twocolumn]{article}
\usepackage{times,graphicx,color}
\usepackage{nar}
\usepackage{epsfig}

\begin{document} 

\title{\textbf{\LARGE Structure and function of negative feedback loops at the interface of genetic and
metabolic networks}}
\author
{Sandeep Krishna, Anna M. C. Andersson, Szabolcs Semsey, Kim Sneppen\footnote{corresponding author: Kim Sneppen: sneppen@nbi.dk}\\
Niels Bohr Institute, Blegdamsvej 17, 2100 Copenhagen, Denmark.}
\date{}

\maketitle 

\section{abstract}
\\ {\bf The molecular network in an organism consists of
transcription/translation regulation, protein-protein
interactions/modifications and a metabolic network, together 
forming a system that allows the cell to respond sensibly to
the multiple signal molecules that exist in its environment.
A key part of this overall system of molecular
regulation is therefore the interface between the genetic
and the metabolic network. 
A motif that occurs very often at this interface is a negative 
feedback loop used to regulate the level of the signal molecules.
In this work we use mathematical models to
investigate the steady state and dynamical behaviour
of different negative feedback loops.
We show, in particular, that feedback loops where the signal molecule
does not cause the dissociation of
the transcription factor from the DNA respond faster than 
loops where the molecule 
acts by sequestering transcription factors off the DNA. 
We use three examples, the {\it bet}, {\it mer}
and {\it lac} systems in {\it E. coli}, to illustrate the 
behaviour of such feedback loops.}

\section{Introduction}\\
Gene expression in bacterial cells is modulated to enhance the cell's
performance in changing environmental conditions. To this end, transcription
regulatory networks continuously sense a set of signals and perform
computations to adjust the gene expression profile of the cell. A
subset of such signals contains molecules that the cell can
metabolize. These molecules range from nutrients to toxic compounds.
A commonly occurring motif in the networks sensing such signal
molecules is a negative feedback loop.  In this motif an
enzyme used to metabolize the signal molecule is controlled by a
regulator whose action, in turn, is regulated by the same signal
molecule. This motif allows for genes that are not transcription
factors to negatively regulate their own synthesis.

Because these negative feedback loops are situated at the
interface of genetic  
\cite{RegulonDB,EcoCyc} and metabolic \cite{EcoCyc,EdwardsPalsson} networks, understanding their
behavior is crucial for building integrated network models, as
well as synthetic gene circuits \cite{GCC,Kobayashietal,Guidoetal}. 
In fact, if one ignores the interface,
the network topology gives the impression that feed back
mechanisms are less frequent than feed forward loops \cite{SMMA,ManganAlon}. 
In addition, by ignoring feedback associated to
signal molecules one would also tend to overemphasize the modular
features of the overall system \cite{HHLM} and underemphasize the average
number of incoming links to proteins. 

\begin{figure}[t]
\epsfig{height=9cm,angle=270,file=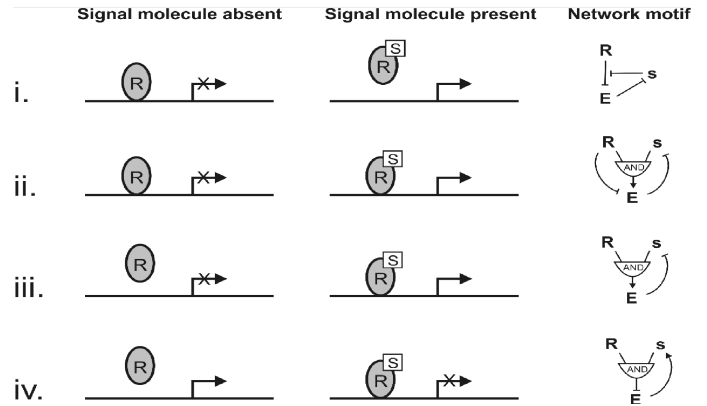}
\caption{\small\sl \label{schematic}Schematic diagrams of four types of negative feedback
loops containing proteins and signal molecules. Regulation of the promoter (arrow)
transcribing the enzyme (E) that can metabolize the signal molecule (s) is shown in
the first two columns. R represents the regulator. A network representation of
each regulation mechanism is shown in the rightmost column. This is not a complete logical
list of all combinatorial possibilities, it includes only the ones where 
real examples have been found.}
\end{figure}

Even within the framework of a negative feedback loop there are several different
mechanisms possible both for transcriptional regulation and for the action
of the signal molecule. We list below four mechanisms which are present
in living cells, with examples taken from {\it E. coli}
(Fig. \ref{schematic}):\\

\noindent
(i) The regulator, R, represses the transcription of the enzyme,
E, which metabolizes the signal molecule, s. The signal molecule
binds to the repressor resulting in the dissociation of the
R-operator complex and an increase
in the production of E. This mechanism 
is exemplified by a negative feedback loop in the {\it lac} system \cite{JacobMonod}, where the roles of R, E and s are played by 
LacI, $\beta$-galactosidase, and lactose, respectively.\\ 

\noindent
(ii) R represses the transcription of 
E which metabolizes s. But here the signal molecule
can bind to R even
when it is at the operator site. When this happens the effect of R 
on the promoter activity is
cancelled, or even reversed.
Two examples of this kind are the {\it bet} \cite{LRetal,RLetal}
and {\it mer} \cite{ABetal} systems, which are involved in the response of cells to
the harmful conditions of osmotic stress and presence of mercury ions, respectively.\\ 

\noindent
(iii) Here the regulator, R,
is an activator of the transcription of E when s
is bound to it. Without the signal molecule, R cannot bind to the 
DNA site and activate transcription.
For instance, MalT in complex with maltose is a transcriptional
activator of genes which metabolize maltose.
This mechanism differs from (ii) in that
in the absence of s, R is a repressor in (ii) while
here it does not affect the promoter activity.\\

\noindent
(iv) Here too, R alone cannot bind to the operator site.
However, in contrast to (iii), R bound to s represses the
transcription of E. 
Further, in this case E increases the production of the signal molecule,
rather than metabolizing it, thereby again making the overall feedback negative.
One such example is the regulation of de novo purine nucleotide
biosynthesis by PurR \cite{MKetal,MengNygaard}.\\

A major difference between these four loops is the manner in
which the signal molecule acts. In (i) 
the binding of s to R drastically reduces its affinity
to the DNA site.
On the other hand, in (ii), (iii) and (iv), the signal molecule 
increases, or does not significantly alter, the binding affinity of R
and can also affect the action of the regulator when it
is bound to the DNA.
Henceforth we will refer to these two methods of action as `mechanism (1)' and `mechanism (2)'.

In this paper we have investigated how this difference
in the mechanism of action of the signal molecule translates
to differences in the steady state and dynamical behaviour of
the simplest kind of negative feedback loops containing proteins and
signal molecules.
These loops have only one step, E,
between the regulator and the signal molecule. Further, the regulator is assumed to
have only one binding site on the DNA. 
We concentrate on
the cases where R is a repressor and s lifts the repression (i and ii).
In particular, we show
that the two mechanisms differ substantially in their dynamic behaviour 
when R is large enough
to fully repress the promoter of E in the absence of s. 
We illustrate how the difference is used in cells by
the examples of the {\it bet}, {\it mer} and {\it lac} systems.

\section{Results}\\
\subsection{Steady state behaviour of feedback loops}\\
First we consider how the steady state activity of the promoter of E responds to changes in
the concentration of s for each of the mechanisms.

Consider a feedback loop, like Fig. \ref{schematic} (i),
where the operator can be found in one of two states: free, $O_{\rm free}$, and 
bound to the regulator, $(RO)$,
with the total concentration of operator sites
being a constant: $O_{\rm tot}=O_{free}+(RO)$.
We assume that the promoter is
active only when the operator is free, and completely repressed when
it is bound by R.
This loop uses mechanism (1) and is an idealization of the {\it lac} system in {\it E. coli}.
The promoter activity is given by:
\begin{equation}
\label{eq:mech1_partial}
A_1=\frac{O_{\rm free}}{O_{\rm tot}}=\frac{O_{\rm free}}{O_{\rm free}+(RO)}.
\end{equation}
In steady state, $O_{\rm free}$ and $(RO)$ can be expressed as functions
of the total concentration of regulators, $R_{\rm tot}$, and the concentration of signal
molecules, $s$. The expression also contains the parameters $K_{RO}$ and $h$ (the equilibrium
binding constant for R-operator binding and the corresponding Hill coefficient), $K_{Rs}$ and
$h_s$ (for R-s binding.) Equation \ref{eq:mech1_full} in the Methods section contains
all the details. The main effect of s is to decrease the amount
of free R because $R_{\rm tot}=R_{\rm free}+(Rs)$, where $(Rs)$ is the concentration
of the R-s complex.

For a feedback loop using mechanism (2),
the operator can be found in one of three states:
free, $O_{\rm free}$, 
bound to the regulator, $(RO)$,
and bound to the regulator along with the signal molecule, $(RsO)$.
The total concentration 
of operator sites, $O_{\rm tot}=O_{free}+(RO)+(RsO)$, is constant.
The promoter activity has a basal value (normalized to 1) when the operator is free.
When the regulator alone is bound it represses the activity. 
We assume the
activity in this state is zero. 
When the operator is bound by R along with s the activity returns to the basal level.
This is an idealization of the {\it bet} system.
Here, the promoter activity is given by:
\begin{equation}
\label{eq:mech2_partial}
A_2=\frac{O_{\rm free}+(RsO)}{O_{\rm free}+(RO)+(RsO)}.
\end{equation}

The main effect of s comes from the second term in the numerator
of equation 2, which is the concentration of the R-s-operator complex.
As in the case of $A_1$, in steady state the activity can be expressed in terms of $R_{\rm tot}$ and $s$.
Because of the third state of the operator, $(RsO)$, the expression for
$A_2$ includes one more parameter, $K_{RsO}$, the equilibirum binding constant
for R-operator binding when s is bound to R (see equation \ref{eq:mech2_full} in the Methods section for details.)

For mechanism (2), we mainly consider the case where $K_{RsO}=K_{RO}$, i.e., the binding of the signal molecule
does not change the binding affinity of the regulator to the operator.
This is the simplest situation and illustrates the basic differences between the two
mechanisms.
In real systems these binding
constants are often different. However, as we show,
for {\it bet} and {\it mer} the inequality of
$K_{RsO}$ and $K_{RO}$ does not obscure the differences caused
by the two mechanisms of action of the signal molecule.
This is because the main effect of changing $K_{RsO}$ is simply to shift the
position of the response curve. Only 
when $K_{RsO}$ becomes very large (which results in 
dissociation of R from the operator when s binds to it, as in {\it lac}) does mechanism (2) effectively
reduce to mechanism (1).

\begin{figure}[t]
\begin{center}
\epsfig{height=8cm,angle=270,file=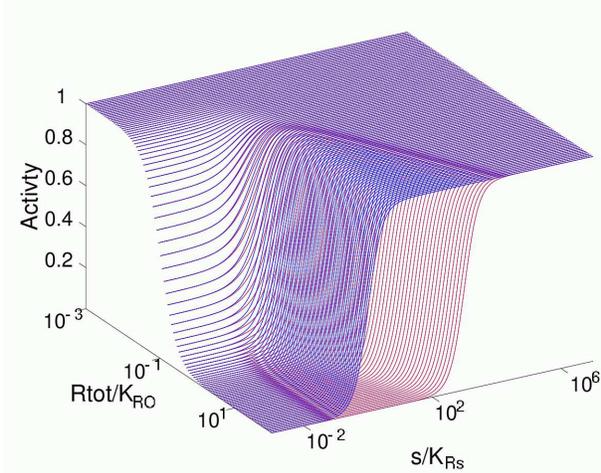}
\end{center}
\vspace{-0.6cm}
\caption{\sl\small 
\label{3dplot}Promoter activities for the two mechanisms as a function of
$R_{\rm tot}$ and $s$. Red: $A_1$ (equation \ref{eq:mech1_partial} and \ref{eq:mech1_full}); Blue: $A_2$ (equation \ref{eq:mech2_partial} and \ref{eq:mech2_full}.)
Parameter values (see equations \ref{eq:mech1_full} and \ref{eq:mech2_full}) are:
$K_{RO}=1, K_{Rs}=100, h=2, h_s=1$ and $K_{RsO}=1$ for mechanism (2).
Choosing $h=2$ assumes that two protein subunits are involved in the R-operator binding.
}
\end{figure}

Figure \ref{3dplot} shows the activities $A_1$ and $A_2$ for a range of values of
$R_{\rm tot}$ and $s$. 
The following observations can be made from the figure:
\begin{itemize}
\item[(a)] For sufficiently small values of $R_{\rm tot}$ there is no difference
between $A_1$ and $A_2$.
\item[(b)] From $R_{\rm tot}/K_{RO}=1$ and higher, $A_1$ requires larger and larger s
to rise to its maximum value, i.e., its effective binding constant $K_{\rm eff}$
increases with $R_{\rm tot}$ (where we define $K_{\rm eff}$ to be the value of $s$
at which the activity is half-maximum.)
\item[(c)] $A_2$, on the other hand, has a $K_{\rm eff}$ which is remarkably
robust to changes in $R_{\rm tot}$, remaining close to $K_{Rs}$ for $R_{\rm tot}/K_{RO}>1$.
\item[(d)] Zooming in to the low $s$ region shows that $A_2$ rises more steeply than $A_1$ for small $s$ values.

\end{itemize}

All these features can be explained by taking a closer look at the equations
for $A_1$ (eq. \ref{eq:mech1_partial} and \ref{eq:mech1_full}) and $A_2$ 
(eq. \ref{eq:mech2_partial} and \ref{eq:mech2_full}.)
Taking the observations in reverse order, first we see that for
small values of $s$, the promoter activities rise as a power of $s$:
$A(s)\approx A(0)+{\rm const.}\times s^a$.
From equations \ref{eq:mech1_full} and \ref{eq:mech2_full} we find that 
this power $a=h_s$ for mechanism (1) and $a=h\times h_s$ for mechanism (2).
Thus, as long as $h>1$, mechanism (2) will have a steeper response at small
values of $s$.

Next, let us consider the amount of inducer $\; s=K_{\rm eff}$
needed to half-activate the promoter under the two mechanisms. The
fact that $K_{\rm eff}$ is close to $K_{Rs}$ for $A_2$ is because of
the term $(RsO)$, which occurs in both the numerator and the
denominator of equation \ref{eq:mech2_partial}. When $R_{\rm tot}$
is large enough (i.e., $R_{\rm tot}>K_{R0}$), the operator is
rarely free, and the constant term (=1) in eq. \ref{eq:mech2_full}
can be disregarded from both numerator and denominator. In that case
$A_2$ only depends on the ratio between the binding affinities $K_{RsO}/K_{RO}$.
Accordingly $A_2$ becomes independent
of the value of $R_{tot}$ for $R_{tot}>K_{RO}=1$.

On the other hand, the activity $A_1$ is always highly dependent on
$R_{tot}$. From equation \ref{eq:mech1_full} we see that $A_1$
reaches half-maximum when $(RO) = O_{\rm free}$. This
happens when $(s/K_{Rs})^{h_s} \approx (R_{\rm tot}/K_{RO})$.
Therefore $K_{\rm eff}$ is an increasing function of $R_{\rm tot}$ 
for mechanism (1).

For both mechanisms, when $R_{\rm tot}$ drops below
$min(K_{RsO},K_{RO})$ we enter a regime where the inducer is not
needed for derepression. For our standard parameters, repressor
concentration $R_{\rm tot}<K_{RO}$ implies that $O_{\rm free}$ and
$(RO)$ dominate $(RsO)$ in equation \ref{eq:mech2_partial}. Thereby
the functional form of the activity $A_2$ approaches that of $A_1$,
as indeed seen from the $R_{\rm tot}/K_{RO}<1$ regime in Fig. 2.

In addition to these mathematical arguments, the above observations
can be understood physically from the nature of the processes
allowed in mechanisms (1) and (2).
Consider the case of a fully repressed promoter (when $R_{\rm
tot}\gg K_{RO}$). Mechanism (1) then requires dissociation of $R$
from the operator for the activity to rise and this is associated
with a free energy cost proportional to ${\rm ln}(R_{\rm
tot}/K_{RO})$. In mechanism (2) there is no such cost and therefore
a smaller amount of $s$ is required to achieve the same level of
inhibition of R. Thus, for genes which are typically completely
repressed, and transcription of which, on the other hand, 
may be needed suddenly,
mechanism (1) is inferior to mechanism (2) because it needs a larger
amount of $s$.
After first discussing three real systems, we will elaborate on this
response advantage by comparing the explicit time dependence of the
two mechanisms in the next section.

The most general framework within which the promoter activities of the enzymes in the
{\it bet}, {\it mer} and {\it lac} systems can be represented is the following generalization
of equations 1 and 2:

\begin{equation}
\label{eq:general_partial}
A=\frac{\alpha O_{\rm free}+\beta (RO)+\gamma (RsO)}{O_{\rm free}+(RO)+(RsO)}.
\end{equation}

Equation \ref{eq:general_full} in the Methods section shows the dependence of $A$ on $R_{\rm tot}$ and $s$.
$\alpha,\beta,\gamma$ are constants dependent on which system we are
trying to describe. 
$\alpha$ is the promoter activity in the absence of R and is used as a reference (1.00).
$\beta$ and $\gamma$ are the relative promoter activities in the presence of R alone,
and R together with s, respectively.
Table 1 shows the values of $\alpha,\beta,\gamma$ 
as well as how the binding affinity of R to the operator is changed by
the binding of s (the ratio $K_{RsO}/K_{RO}$) for the three
systems. We have used the Hill coefficients 
$h=2$ (assuming that two protein subunits are involved in DNA binding) and $h_s=1$ (for simplicity and to
compare with Fig. \ref{3dplot}.)
Mechanism (1) and (2) are special cases of this equation. Equations \ref{eq:mech1_partial} and \ref{eq:mech1_full},
for mechanism (1), are obtained by setting $\alpha=1,\;\beta=0$ and taking the limit $K_{RsO}\rightarrow\infty$ (the value of $\gamma$ is irrelevant in this limit).
Equations \ref{eq:mech2_partial} and \ref{eq:mech2_full},
for mechanism (2), are obtained by setting $\alpha=1,\;\beta=0,\;\gamma=1$.
From Table 1, it is clear then that {\it lac} uses mechanism (1) and
{\it bet} uses mechanism (2). {\it mer} is an even more extreme case of
mechanism (2) where the $(RsO)$ term has a much larger weight ($\gamma\gg 1$)
than the idealized mechanism (2).

\begin{table}
\begin{tabular}{|c|c|c|c|c|c|}
\hline
System & $\alpha$ & $\beta$ & $\gamma$ & $K_{RsO}/K_{RO}$ & References\\
\hline
bet & 1 & 0.32 & 0.83 & 0.29 &\cite{LRetal,RLetal}\\
mer & 1& 0.13 & 13.71 & 3.5 &\cite{ABetal}\\
lac & 1& 0.06 & 1& 1000&\cite{OEetal,BRetal}\\
\hline
\end{tabular}
\caption{Values of parameters in equation 3 for three systems found in {\it E. coli}.
In the case of {\it lac} we used a simplified case, where the {\it lac} promoter is repressed by
LacI binding to a single operator, {\it O1}.
}
\end{table}

\begin{figure}[ht]
\begin{center}
\epsfig{width=8cm,file=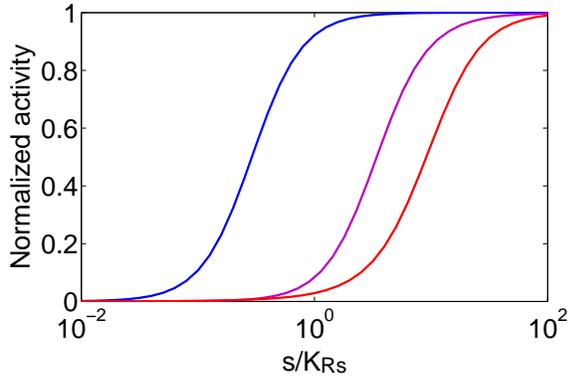}
\end{center}
\vspace{-0.6cm}
\caption{\sl\small \label{4systems}Promoter activity, $A$, (see equation \ref{eq:general_partial} and \ref{eq:general_full}) versus $s$ for the {\it bet} (blue), 
{\it mer} (purple),
{\it lac} (red) systems in {\it E. coli}. Because the promoter activities for the three systems have different
minimum and maximum values, for easier comparison we have plotted $(A-A_{\rm min})/(A_{\rm max}-A_{\rm min})$,
where $A_{\rm min}=A(s=0)$ and $A_{\rm max}=A(s=\infty)$. 
In all cases, we kept $R_{tot}/K_{RO}=10$.}
\end{figure}

Fig. \ref{4systems} shows the response curves for the {\it bet}, {\it mer} and {\it lac} systems.
{\it bet} and {\it mer}, representatives of mechanism (2), and {\it lac}, a representative of
mechanism (1), indeed behave similar to the idealized versions of
the two mechanisms investigated in Fig. \ref{3dplot}.
The difference between {\it bet} and {\it mer} is the result of changes in the binding
affinity of R in the absence and presence of s.

A further complication that could occur in real systems is that the probabilities
of RNA polymerase recruitment could be different for different states of the operator.
We find that taking the changing probabilities of RNA polymerase recruitment
into account does not change the mathematical form of the equations for
the promoter activities (see Methods.) Thus, this additional complication does not affect
our results. 

\subsection{Dynamical behaviour of feedback loops}\\
We now turn to an analysis of differences in the temporal behaviour of
the feedback mechanisms. We model the dynamics
by two coupled differential equations:

$$\frac{dE}{dt}=A(R_{\rm tot},s)-E$$
and
$$\frac{ds}{dt}=c-kEs.$$

$E$ and $s$ represent the concentrations of the enzyme and signal molecule, respectively.
The first term in the $dE/dt$ equation is the rate of production of E which
is equal to the promoter activity, $A$ (equations \ref{eq:general_partial} and 
\ref{eq:general_full})\footnote{In using the steady state expressions
for activities we are, in effect, assuming that the binding and dissociation of R to the operator and s to R occur on a much
faster timescale than the transcription and translation of E.}. The second term
represents degradation of E. 
The second equation describes the evolution of the concentration of s;
it increases if there is a source, $c>0$, of s
(for instance from outside the cell) and decreases due to the action of the enzyme E.
In the first equation both terms could be multiplied by rate constants, representing
the rates of transcription, translation and degradation. However, we have eliminated
these constants by measuring time, $t$, in units of the degradation time of E, and
by rescaling $E$ appropriately (see the Methods section for details.)
Thus, in these equations, $E$ and $t$ are dimensionless, with $E$ lying between 0 and 1.
$k$ can then be interpreted as the maximum rate of degradation of s in units of the degradation
rate of E.

\begin{figure}[th!]
{\bf a.\\}
\epsfig{width=4.5cm,file=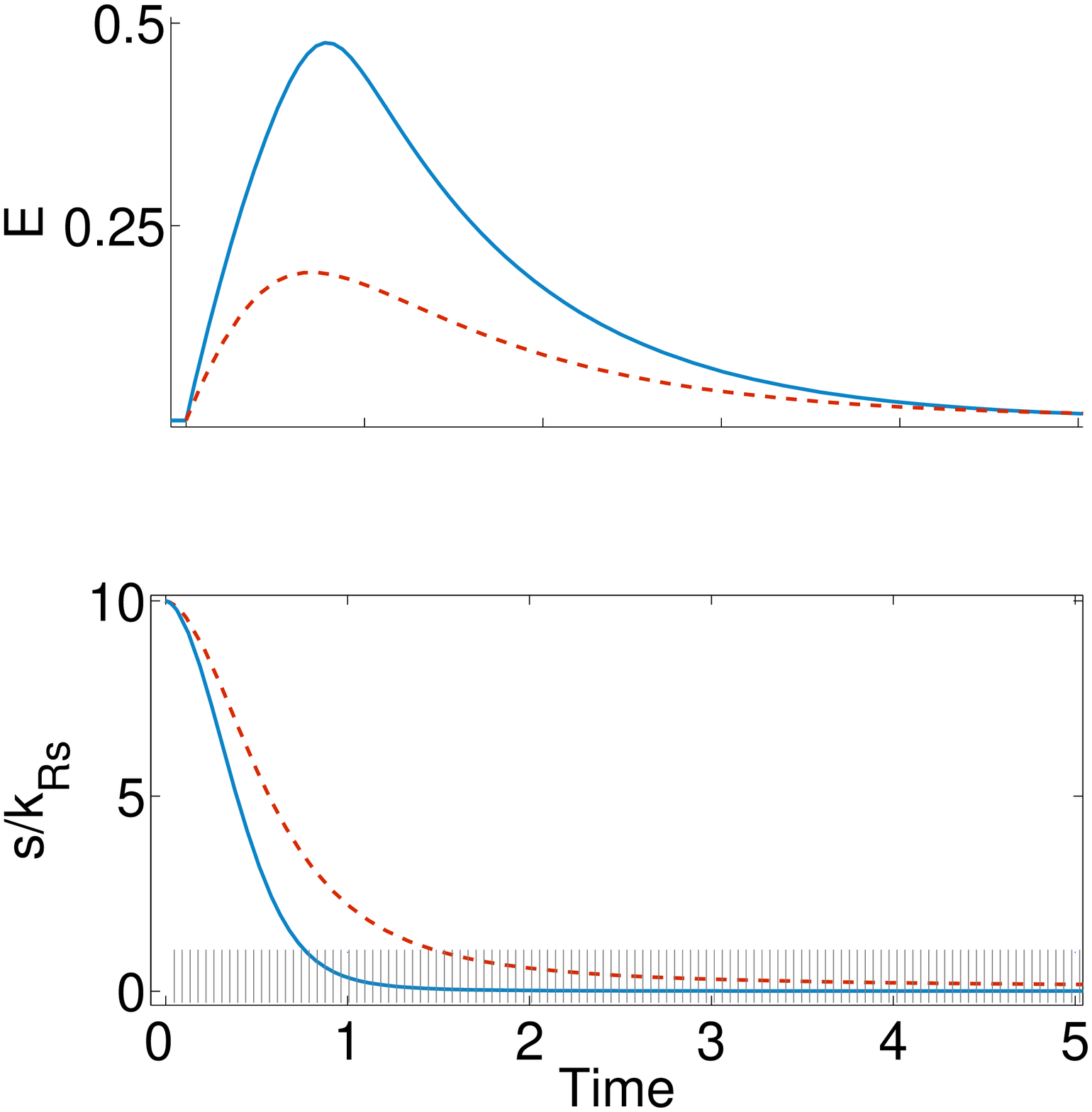}
\epsfig{width=4.5cm,file=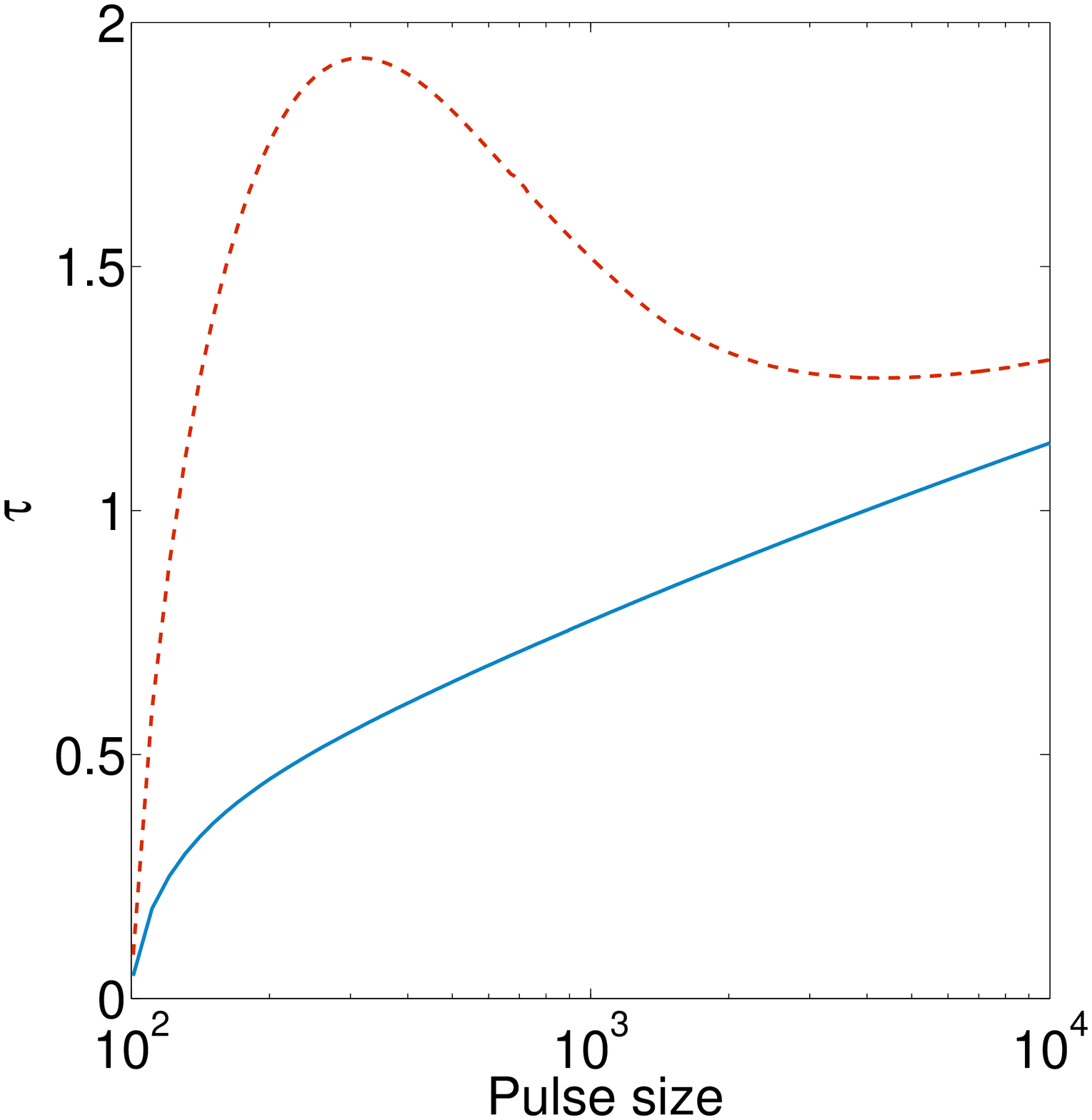}\\
{\bf b.\\}
\epsfig{width=4.5cm,file=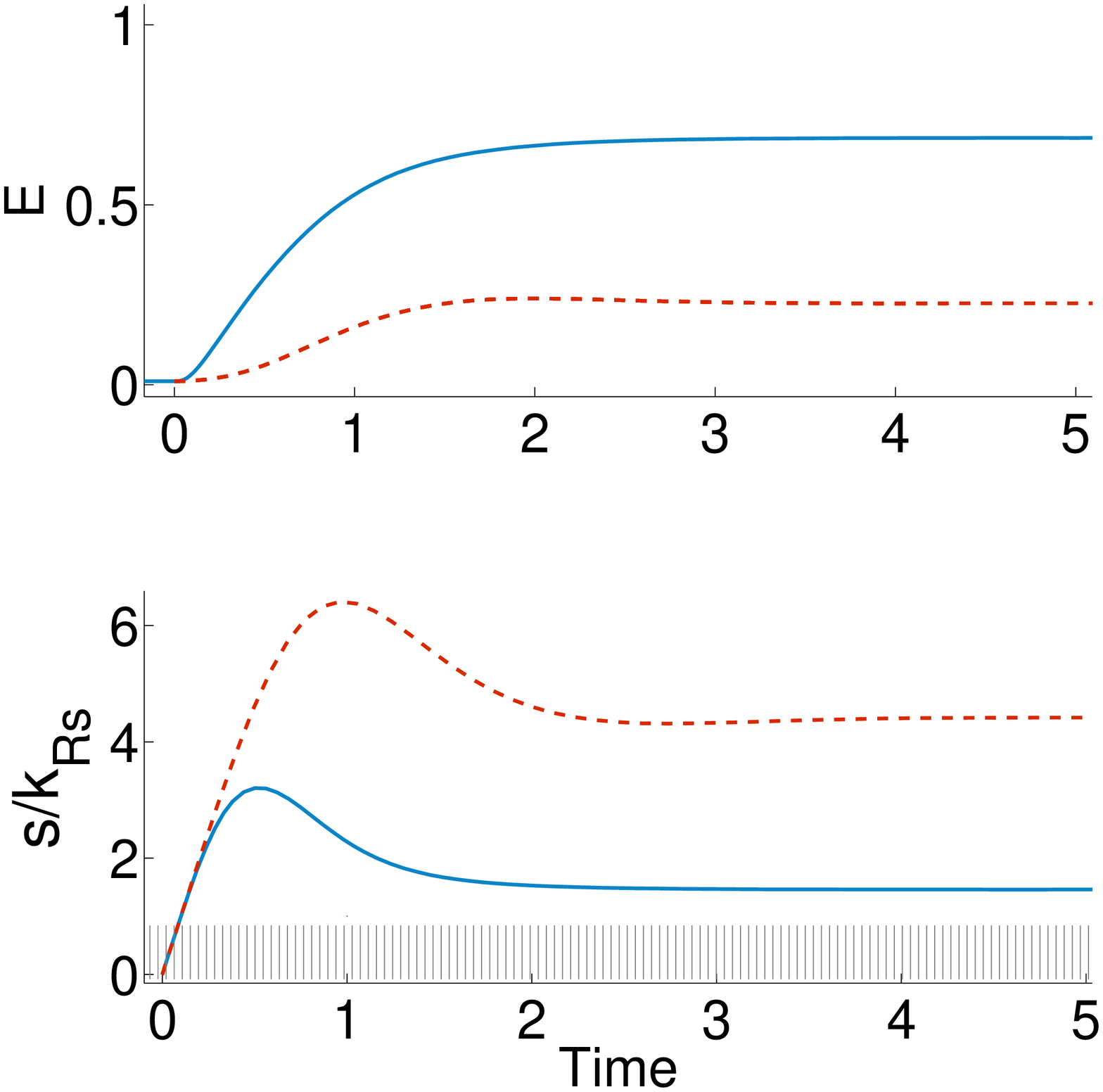}
\epsfig{width=4.5cm,file=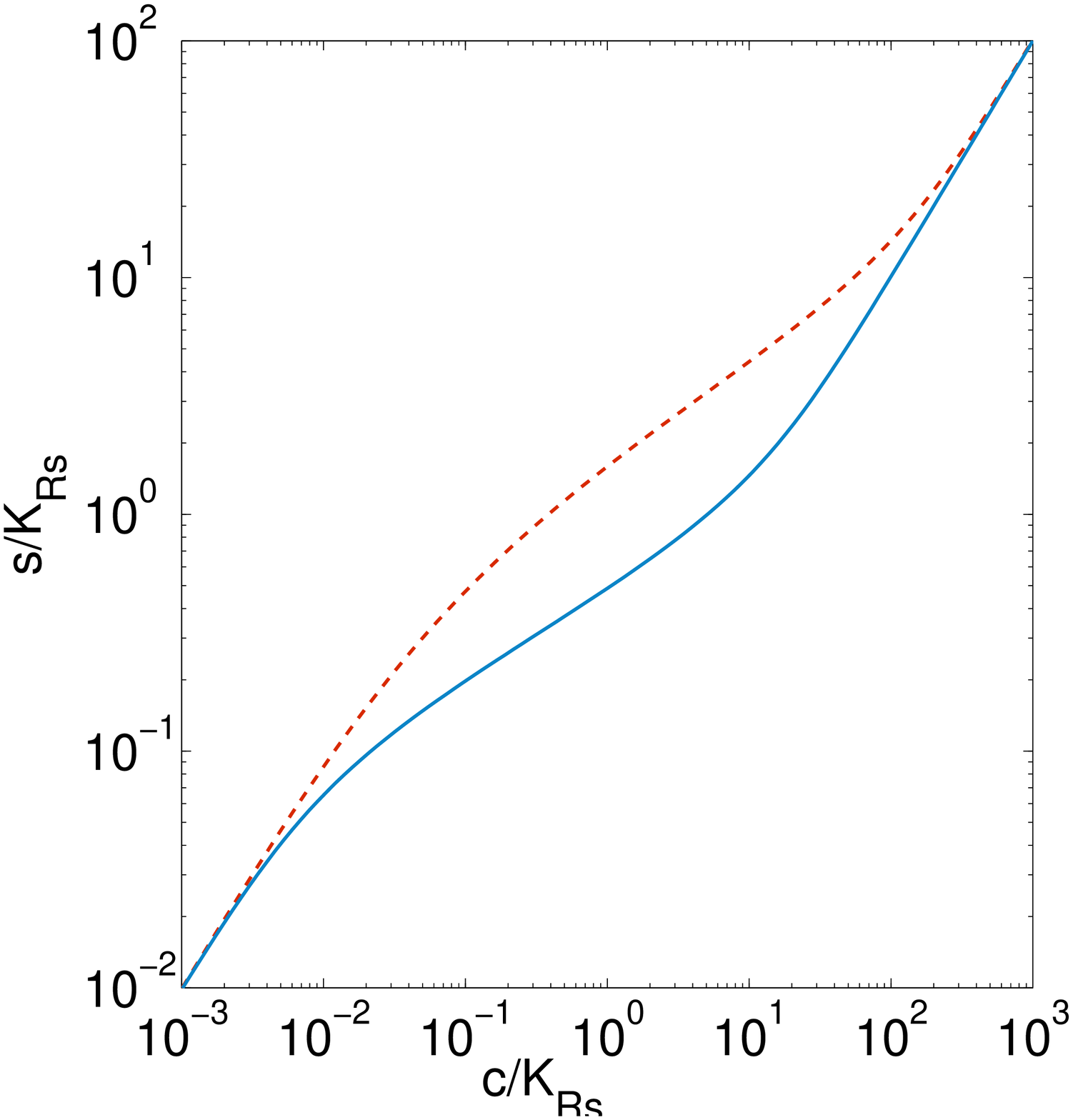}
\caption{\small\sl \label{dynamics}{\bf a.} Left panel: time course for 
levels of E and s when there is no source of s ($c=0$) and the system
is subject to an instantaneous pulse lifting the value of $s$ from 
zero to $10\times K_{Rs}$ at time $t=0$. The shaded region corresponds to $s<K_{Rs}$. 
Right panel: the time (after the pulse), $\tau$, 
required for each mechanism to reduce the value of $s$ down to $K_{Rs}$,
for different sizes of the initial pulse. 
{\bf b.} Left panel: time course of levels of E and s when a source of s 
($c=10\times K_{Rs}$ per degradation time of E) 
is suddenly introduced at time $t=0$. 
Right panel: Difference between the steady state values of 
$s$ for the two mechanisms as a function of the value of 
the source term, $c$. 
In all plots the red, dashed line indicates mechanism (1) and the blue, 
solid line, mechanism (2).
Parameter values are: $K_{RO}=1, K_{Rs}=100, h=2, h_s=1, k=10$, 
and $K_{RsO}=1$ for mechanism (2).
The active degradation rate $k$ has been chosen to be much
larger than the degradation rate of E (which is unity).}
\end{figure}

Fig. \ref{dynamics}a (left panel) shows what happens if the cell is subject to a sudden
pulse of s. That is, the source $c=0$ always, but at time $t=0$ the concentration of s
abruptly jumps from zero to $10\times K_{Rs}$. This triggers an increase in the production of E
which then starts to decrease the concentration of s. There is no further
addition of s to the system, so eventually all of it is removed and
the system returns to its condition before the pulse. From the
figure we see that, for the same parameter values, 
mechanism (2) results in a much faster removal of s
because the response of E to the pulse is larger.
The right panel adds further evidence to this conclusion. It shows, for both
mechanisms, perturbed by varying sized pulses of s, the time taken for the
concentration of s to fall to $K_{Rs}$.
This measure shows that mechanism (2) generally responds
faster than mechanism (1). 
The two mechanisms converge for small perturbations because
there is no signal to respond to (levels of $s$ are very low),
and for very large perturbations because then the 
promoter becomes fully activated by the huge concentration  
of the inducing molecule s.

Fig. \ref{dynamics}b shows what happens when the cell is subject to the
appearance of a constant source of s. At time $t=0$ the value of $c$
abruptly jumps from zero to $10\times K_{Rs}$ per degradation
time of E. In response, the production of E is increased
and eventually reaches a new steady state value to deal with the constant
influx of s (left panel). From the right panel of the figure 
it is evident that mechanism (2) is able to
suppress the amount of s much more than mechanism (1) for most values 
of the rate of influx. Again, for similar reasons, the two mechanisms
converge at small and large values of $c$.

These observations apply for the case when $K_{RsO}=K_{RO}$ for mechanism (2). 
The only effect on
the dynamical equations caused by changing the ratio $K_{RsO}/K_{RO}$ lies
in the expression for $A$ in the first term of the $dE/dt$ equation. As mentioned
in the previous section, changing this ratio mainly results in shifting of the response
curve and as $K_{RsO}$ is increased, $A_2$ approaches $A_1$. For the dynamics
this results in an increase in $\tau$ (for a pulse) and in the steady state value
of $s$ (for a source) as $K_{RsO}/K_{RO}$ is increased. These values approach
those for mechanism (1) in the limit $K_{RsO}/K_{RO}\rightarrow\infty$. The amount
by which $K_{RsO}$ has to be boosted to effectively reduce mechanism (2) to
mechanism (1) increases with increasing $R_{\rm tot}$, as in the steady state case.
 
\section{Discussion}\\

In the present paper we have discussed various strategies for
negative feedback mechanisms involving the action of one 
signal molecule on a transcription factor. In particular, we have
investigated two broadly different ways in which the signal molecule may change
the action of the transcription factor: first, it could inhibit its action 
by sequestering it, and second, it could bind to the transcription factor
while it is on the DNA site and there alter its action.
The first mechanism occurs when the binding of the signal molecule
reduces the affinity of the transcription factor 
to such an extent that it cannot subsequently remain bound to the DNA.
This kind of inhibition of the transcription factor 
occurs in the {\it lac} system, where (allo)lactose reduces the
binding affinity of LacI to the operator {\it O1} by a factor of 1000 \cite{BRetal}.
This mechanism has also been exploited in synthetic gene networks \cite{GCC,Kobayashietal}.
In the second mechanism the binding affinity is not altered that much;
{\it bet} and {\it mer} belong to this category. In the case of {\it mer}
the presence of the signal molecule reverses the action of the transcription
factor, changing it from a repressor to an activator \cite{ABetal}.

In steady state, the two mechanisms differ most when the levels of the
transcription factor are large enough to ensure substantial repression
in the absence of the signal molecule. 
The underlying reason for these differences is that, in this regime of full repression,
for each transcription factor that
binds to the signal molecule there is, for mechanism (1), an extra energy cost 
for the dissociation of the transcription factor from the DNA.
The dynamical behaviour of feedback loops based on mechanisms (1) and (2)
also differ substantially when promoters are in the fully repressed regime.
We have shown that when the systems are perturbed by the sudden appearance
of either a pulse or a source of signal molecules, mechanism (2) is generally
faster and more efficient than mechanism (1) in suppressing the levels of the molecule.
This prediction could be tested using synthetic gene circuits
which implement these two mechanisms, for instance by extending the circuits
built in ref. \cite{Guidoetal}.
In addition, this observation 
fits neatly with the fact 
that the {\it bet} and {\it mer} systems use versions of mechanism (2), because
they respond to harmful conditions (osmotic stress and the presence of mercury ions, respectively)
and therefore need to respond quickly, while mechanism (1) is associated with {\it lac},
a system involved in metabolism of food molecules which therefore
does not need to be as sensitive to the concentration of the signal molecules.
In the case of {\it lac} it is probably energetically disadvantageous for the cell
to respond to low levels of lactose sources \cite{DA}.

The differences between mechanisms are clear when they are compared
keeping all parameters constant. In cells, however, parameter values
vary widely from one system to another which can obscure the differences
caused by the two mechanisms. For instance, it is possible to increase the
speed of response of mechanism (1) by reducing the $K_{Rs}$ value (i.e.,
increasing the binding strength between the regulator and the signal molecule.)
Keeping all other parameters constant $K_{Rs}$ needs to be decreased by a factor
10 for mechanism (1) to behave the same as mechanism (2) when $R_{\rm tot}=10\times K_{RO}$. 
This factor increases as $R_{\rm tot}$ increases, i.e., as the repression is more complete.
This can again be understood
in terms of the extra energy cost for mechanism (1): increasing $K_{Rs}$ sufficiently
makes the extra energy cost insignificant compared to the R-s binding
energy. Thus, a negative feedback loop in a real cell which needs to respond to
signals on a given fast timescale could do so either by using mechanism (2), or
by using mechanism (1) with a substantially larger R-s binding affinity.
For signal molecules where it is not possible for the R-s binding to be
arbitrarily strengthened, mechanism (2) would be the better choice.
On the other hand, mechanism (2) also has its disadvantages.
For instance, at promoters with complex regulation the DNA bound
transcription factor using mechanism (2) may interfere with the action 
of other transcription factors.

In Figure 1 we showed 4 examples, and have extensively discussed example (i) and (ii).
Another implementation of mechanism (2) is example (iii), with an activity 
$A_3=(RsO)/[O_{\rm free}+(RsO)]$. In general this regulatory module is at least as efficient as
mechanism (2), with a dynamical response which is even more efficient
in the intermediate range of $s$ (around $K_{Rs}$). The loop
in Fig. \ref{schematic}(iv) is, on the other hand, a different kind of negative feedback
from the other three examples. It involves synthesis of the signal molecule, and thus
is aimed at maintaining a certain concentration of the molecule, rather then
minimising or consuming it.
In practice, it is the kind of feedback that is common in biosynthesis pathways, 
where it helps maintain
a certain level of amino acids, nucleotides, etc., inside the cell.

The simple one-step, single-operator negative feedback loops investigated
here clearly indicate that the mechanism of action of the signal molecule
is a major determinant of the steady state and dynamical behaviour of
the loop. Additional complexity in the mechanism of regulation
(e.g., cooperative binding of a
transcription factor to multiple binding sites) or of the regulatory
region (competing transcription factors or multiple regulators responding to
different signals) \cite{BBetal1,BBetal2} would open up more avenues for the differences between the two
mechanisms to manifest themselves.

These feedback loops form the link connecting the
genetic and metabolic networks in cells. In fact, such loops involving signal molecules
are likely to be a dominant mechanism of feedback regulation of transcription.
Feedback using only regulatory proteins, without signal molecules, is probably too
slow because it relies on transcription to change the levels of the proteins. 
Negative auto-regulation can speed up the response of transcription regulation \cite{REA}. Nevertheless,
feedback loops based on translation regulation \cite{Gottesman,AxelsenSneppen}, active protein degradation
\cite{GZZBG,APS} or 
metabolism of signal molecules
will certainly be able to operate on much faster timescales.
This is probably why feedback loops are rare in purely transciptional networks,
which has contributed to the view that feed forward loops are dominant motifs
in transcription regulation. 
Taking feedback loops involving signal molecules 
into account alters this viewpoint substantially. 
In  {\it E. coli} the number of feedforward loops in the transcription regulatory network
has been reported to be 40 \cite{SMMA,ManganAlon}. Based on data in the EcoCyc database \cite{EcoCyc}, we know
that there are more than 40 negative feedback loops involving signal molecules
where the regulation is by a transcription factor.
Adding this many feedback loops to the genetic network would also change
the network topology substantially. In particular, it would diminish the distinction
between portions of the network that are downstream and upstream of a given protein.
The effect of this would be to make the network more interconnected
and reduce the modularity of the network by increasing the number of links between apparently
separate modules. 

\section{Methods}\\
\subsection{Promoter activity for mechanism (1)}\\
The operator can exist in one of two states: (i) free, $O_{\rm free}$,
and (ii) bound to the regulator, $(RO)$.
If the concentration of free regulators is $R_{\rm free}$ then
$$(RO)=\left(\frac{R_{\rm free}}{K_{RO}}\right)^h O_{\rm free}.$$
Similarly, the concentration of regulators bound to signal molecules is
$$(Rs)=\left(\frac{s}{K_{Rs}}\right)^{h_s} R$$
and the total concentration of regulators, a constant, is given by
$$R_{\rm tot}=R_{\rm free} + (RO) + (Rs).$$
We assume that the number of signal molecules is much larger than the number
of regulators which, in turn, is much larger than the number of operator sites,
i.e., $s\gg R_{\rm tot}\gg O_{\rm tot}$.
Then we can take $s$ to be approximately constant and we can take $(RO)\ll (Rs)$,
giving:
$$R_{\rm tot}=R_{\rm free}+\left(\frac{s}{K_{Rs}}\right)^{h_s} R_{\rm free},$$
\begin{equation}
\label{eq:R}
\Rightarrow R_{\rm free}=\frac{R_{\rm tot}}{1+(s/K_{Rs})^{h_s}}
\end{equation}
and
\begin{equation}
\label{eq:RO}
(RO)=\left[\frac{(R_{\rm tot}/K_{RO})}{1+(s/K_{Rs})^{h_s}}\right]^h O_{\rm free}
\end{equation}

Using these and $O_{\rm tot}=O_{\rm free} + (RO),$ we get:
\begin{equation}
\label{eq:mech1_full}
A_1\equiv\frac{O_{\rm free}}{O_{\rm tot}}=\frac{1}{1+\left[\frac{(R^{tot}/K_{RO})}{1+(s/K_{Rs})^{h_s}}\right]^h}.
\end{equation}

\subsection{Promoter activity for mechanism (2)}\\
The operator can exist in one of three states: (i) free, $O_{\rm free}$,
(ii) bound to the regulator, $(RO)$, and (iii) bound to the regulator along with the
signal molecule, $(RsO)$.

Again, with similar assumptions, we get equation \ref{eq:R} and \ref{eq:RO}
for $R_{\rm free}$ and $(RO)$ plus an additional expression for $(RsO)$:
\begin{equation}
\label{eq:RsO}
(RsO)=\left[\frac{(R_{\rm tot}/K_{RsO})(s/K_{Rs})^{h_s}}{1+(s/K_{Rs})^{h_s}}\right]^h O_{\rm free}.
\end{equation}

Using $O_{\rm tot}=O_{\rm free} + (RO) + (RsO)$ and equation \ref{eq:mech2_partial} for $A_2$, we get:
\begin{equation}
\label{eq:mech2_full}
A_2=\frac{1+\left[\frac{(R^{tot}/K_{RsO})(s/K_{Rs})^{h_s}}{1+(s/K_{Rs})^{h_s}}\right]^h}{1+\left[\frac{(R^{tot}/K_{RO})}{1+(s/K_{Rs})^{h_s}}\right]^h+\left[\frac{(R^{tot}/K_{RsO})(s/K_{Rs})^{h_s}}{1+(s/K_{Rs})^{h_s}}\right]^h}.
\end{equation}

\subsection{General expression for promoter activity}\\
The most general expression for the activity, shown in equation 3, can also be rewritten
using the expressions for $(RO)$ and $(RsO)$ calculated above:

\begin{equation}
\label{eq:general_full}
A=\frac{\alpha+\beta\left[\frac{(R^{tot}/K_{RO})}{1+(s/K_{Rs})^{h_s}}\right]^h+\gamma\left[\frac{(R^{tot}/K_{RsO})(s/K_{Rs})^{h_s}}{1+(s/K_{Rs})^{h_s}}\right]^h}{1+\left[\frac{(R^{tot}/K_{RO})}{1+(s/K_{Rs})^{h_s}}\right]^h+\left[\frac{(R^{tot}/K_{RsO})(s/K_{Rs})^{h_s}}{1+(s/K_{Rs})^{h_s}}\right]^h}.
\end{equation}

\subsection{Taking into account RNA polymerase recruitment}\\

A more correct, but more cumbersome, way to calculate the promoter activities
is to explicitly take RNA polymerase into account. Then, in the most general case,
the system can be in one of 6 states:

\begin{itemize}
\item[(1)] R not bound to operator, RNAP not recruited: weight=1.
\item[(2)] R not bound to operator, RNAP recruited: wt=$p_1P$.
\item[(3)] R bound to operator, RNAP not recruited: wt=$(RO)$.
\item[(4)] R bound to operator, RNAP recruited: wt=$(RO)p_2P$.
\item[(5)] R-s bound to operator, RNAP not recruited: wt=$(RsO)$.
\item[(6)] R-s bound to operator, RNAP recruited: wt=$(RsO)p_3P$.
\end{itemize}

Here $p_{1,2,3}$ are the probabilities (per concentration) for recruitment of RNA polymerase in the three
different states of the operator, and $P$ is the concentration of RNA polymerase.
Taking the promoter activity to be 0 when the polymerase is not recruited and
$\alpha',\beta',\gamma'$ in states (2), (4) and (6), respectively, the activity can be written as follows:

$$A=\frac{(p_1\alpha'+p_2\beta' (RO)+p_3\gamma' (RsO))P}{1+p_1+(1+p_2)(RO)+(1+p_3)(RsO)}.$$
By absorbing the constants $p_{1,2,3}$ into $K_{Rs}$, $K_{RO}$ and $K_{RsO}$,
we recover equation \ref{eq:general_full}.

\subsection{Rescaling of the dynamical equations}\\

With all rate constants included, the dynamical equations for
the time evolution of the concentrations of E and s can
be written as follows:
$$\frac{dE}{dt}=k_1A(R_{\rm tot},s)-\gamma E,$$
$$\frac{ds}{dt}=c'-k'Es.$$

Now measuring time in units of the degradation time of E: $t'=\gamma t$,
and transforming E using $E'=E(\gamma/k_1)$, we get
$$\frac{dE'}{dt'}=A(R_{\rm tot},s)- E'$$
$$\frac{ds}{dt'}=\frac{c'}{\gamma}-\frac{k'k_1}{\gamma^2}E's$$
which, with $c\equiv c'/\gamma$ and $k\equiv k'k_1/\gamma^2$, are the equations used in the main text.

\section{Acknowledgements}\\
This work was supported by the Danish National Research Foundation.

\end{document}